\documentclass[aps,prd,preprintnumbers,showpacs,twocolumn,groupedaddress,nofootinbib]{revtex4}
\usepackage{graphicx}
\usepackage{latexsym}
\def\beq{\begin{equation}}
\def\eeq{\end{equation}}
\def\bey{\begin{eqnarray}}
\def\eey{\end{eqnarray}}

\def\lsim{\mathrel{\raise.3ex\hbox{$<$\kern-.75em\lower1ex\hbox{$\sim$}}}}
\def\gsim{\mathrel{\raise.3ex\hbox{$>$\kern-.75em\lower1ex\hbox{$\sim$}}}}

\begin{document}

\title{Implications of a 130 GeV Gamma-Ray Line for Dark Matter}  
\author{Matthew R.~Buckley$^1$ and Dan Hooper$^{1,2}$}
\affiliation{$^1$Center for Particle Astrophysics, Fermi National Accelerator Laboratory, Batavia, IL 60510, USA}
\affiliation{$^2$Department of Astronomy and Astrophysics, University of Chicago, Chicago, IL 60637, USA}

\date{\today}

\begin{abstract}

Recent reports of a gamma-ray line feature at $\sim$130 GeV in data from the Fermi Gamma-Ray Space Telescope have generated a great deal of interest in models in which dark matter particles annihilate with a sizable cross section to final states including photons. In this article, we take a model-independent approach, and discuss a number of possibilities for dark matter candidates which could potentially generate such a feature. While we identify several scenarios which could lead to such a gamma-ray line, these models are each fairly constrained. In particular, viable models require large couplings ($g\gsim1-3$), and additional charged particles with masses in the range of approximately $\sim$130-200 GeV. Furthermore, lower energy gamma-ray constraints from the Galactic Center force us to consider scenarios in which the dark matter annihilates in the early universe through velocity-suppressed processes, or to final states which yield relatively few gamma-rays (such as $e^+ e^-$, $\mu^+ \mu^-$, or $\nu\bar{\nu}$). An exception to these conclusions can be found in models in which the dark matter annihilates to heavy intermediate states which decay to photons to generate a line-like gamma-ray spectrum.

\end{abstract}

\pacs{95.35.+d; FERMILAB-PUB-12-259-A}
\maketitle

\section{Introduction}

Dark matter particles which annihilate to final states that include photons, such as $\gamma \gamma$, $\gamma Z$, or $\gamma h$, can provide a mono-energetic gamma-ray feature that is potentially distinguishable from otherwise challenging astrophysical backgrounds. Such gamma-ray lines are generally produced only through loop-level processes, however, suppressing their brightness relative to gamma-rays produced through tree-level processes (for examples of line-generating processes predicted in a variety of dark matter models, see Refs.~\cite{Bergstrom:1997fh,Cline:2012nw,Bertone:2009cb,Ullio:1997ke,bergstromkaplan,Choi:2012ap,Anchordoqui:2009bn,Mambrini:2009ad,Dudas:2009uq,Chalons:2011ia,Jackson:2009kg,Profumo:2010kp,Gustafsson:2007pc,Bertone:2010fn,Acharya:2012dz,Goodman:2010qn,Lee:2012bq,Dudas:2012pb,Kyae:2012vi}).

Recently, there has been a great deal of interest in dark matter annihilations to gamma-ray lines, inspired in large part by the papers by Bringmann {\it et al.}~\cite{Bringmann:2012vr} and Weniger~\cite{Weniger:2012tx}, each of which claim to have identified a tentative spectral feature in the publicly available data of the Fermi Gamma-Ray Space Telescope~\cite{data}. This feature appears at an energy of approximately 130 GeV, and has been detected primarily within regions of the sky which reside within $\sim$20-30$^{\circ}$ of the Galactic Center.

These two paper have considered interpretations of this spectral feature in terms of either dark matter annihilations to charged final states with a gamma-ray contribution from internal bremsstrahlung~\cite{Bringmann:2012vr}, or dark matter annihilations to a mono-energetic $\gamma \gamma$ line~\cite{Bringmann:2012vr,Weniger:2012tx}. While the statistical significance of this feature is modest (3.1$\sigma$ and 3.3$\sigma$ are quoted in the two studies, respectively, each after taking into account the look-elsewhere effect), it could potentially become very statistically significant with additional data from Fermi (and with the improvements anticipated from the release of the Pass 8 version of the Fermi-LAT analysis software~\cite{bubble}). To account for this feature with dark matter particles distributed according to a Navarro-Frenk-White (NFW) profile and annihilating to $\gamma \gamma$, Ref.~\cite{Weniger:2012tx} finds that an annihilation cross section of $\sigma v_{\gamma \gamma} = (2.27 \pm 0.57^{+0.32}_{-0.51}) \times 10^{-27}$ cm$^3$/s is required to this final state. The subsequent analysis of Ref.~\cite{Tempel:2012ey} yields a similar, but somewhat higher, value of $\sigma v_{\gamma \gamma} = (5.1 \pm 2.7) \times 10^{-27}$ cm$^3$/s. The authors of Ref.~\cite{Tempel:2012ey} also find that annihilations to $\gamma \gamma$ provide a considerably better fit to the observed gamma-ray spectrum than that predicted from other final states (with the exception of internal bremsstrahlung and those yielding a narrow box-like spectrum). 

Interpretations of this gamma-ray feature as a signature of dark matter annihilation have not gone unopposed. Shortly after the appearance of Refs.~\cite{Bringmann:2012vr,Weniger:2012tx}, Profumo and Linden~\cite{bubble} pointed out that the regions of the sky identified in those papers as containing this signal are quite similar to those known to be occupied by the so-called Fermi bubbles~\cite{Su:2010qj}. Furthermore, the spectrum of the Fermi bubbles is observed to break in the vicinity of 110-150 GeV, providing a plausible explanation for the appearance of a spectral feature at 130 GeV. This possibility was subsequently addressed in Ref.~\cite{Tempel:2012ey}, which independently analyzed the Fermi data and found that the 130 GeV feature does not correlate with the spatial morphology of the Fermi bubbles (in particular, they find that the signal is brightest near the Galactic Center rather than in the outer bubble regions, as would be expected from annihilating dark matter). In any case, gamma-rays associated with the Fermi bubbles represent a background which are likely to complicate searches for gamma-ray lines in the energy range under consideration. 

Even more recently, the Fermi Collaboration has added their voice to this discussion~\cite{fermi}, placing upper limits on the dark matter's annihilation cross section to mono-energetic lines ($\sigma v_{\gamma \gamma} < 1.4 \times 10^{-27}$ cm$^3$/s for a dark matter mass of 130 GeV, and assuming an NFW profile). And while the Fermi Collaboration does not make any mention of the results of Bringmann {\it et al.} or Weniger in their paper, their new limits appear to be in mild tension with the claims presented in those studies.\footnote{There are several significant differences between the Fermi collaboration line analysis~\cite{fermi} and those of Refs.~\cite{Bringmann:2012vr,Weniger:2012tx}. In particular, while Refs.~\cite{Bringmann:2012vr,Weniger:2012tx} use the up-to-date Pass 7 event selection, the Fermi collaboration's analysis uses Pass 6. The collaboration analysis also considers a larger region of the sky relative to those of Refs.~\cite{Bringmann:2012vr,Weniger:2012tx}, which focus on the regions with the highest expected signal-to-background.} Lastly, the independent analysis of Ref.~\cite{Boyarsky:2012ca} argues that the variability in the spectum observed by Fermi across the sky is quite high, suggesting the presence of significant systematic errors and bringing into question the line-like nature of the gamma-ray feature in question.

In this article, we do not attempt to resolve the questions of the origin (or existence) of this gamma-ray feature, but instead consider the characteristics of dark matter models that are capable of producing such a bright mono-energetic line. In particular, requiring only that such a dark matter candidate does not violate constraints from other indirect detection searches (such as those derived from gamma-ray observations of the Galactic Center), avoids being overproduced in the early Universe, and annihilates through perturbative interactions, we find that only a relatively narrow range of models can lead to such a spectral feature. We discuss each of these possibilities in turn, as well as the relevant constraints and prospects for testing them in other experiments, including at the Large Hadron Collider.

\section{Dark Matter Annihilations to Gamma-Ray Lines}

There are several classes of Feynman diagrams through which two dark matter particles could potentially annihilate to a final state including one or more photons. In this section, we focus on the final state $\gamma \gamma$, and consider four classes of diagrams which could produce the reported line signal. In the first three of these four cases, the photons are produced through diagrams which include a charged particle loop. If the mass of the charged particle(s) is lighter than the mass of the dark matter, then the corresponding tree-level cross section for annihilations to these charged particles will exceed the loop-level process to $\gamma \gamma$ by many orders of magnitude (if the charged particle(s) is only slightly heavier than the dark matter, tree-level annihilations to off-shell pairs of these particles could still potentially dominate). As we will discuss in Sec.~\ref{constraints}, such an enormous annihilation cross section to charged particles is in considerable violation of gamma-ray constraints from observations of the Galactic Center and elsewhere. With this in mind, we are forced to consider processes in which particles heavier than the dark matter itself are exchanged. 

Throughout the remainder of this section, we have adopted specific spin assignments for the dark matter candidate. The conclusions reached, however, are not materially impacted by this these choices.

\subsection{Resonant Diagrams}

\begin{figure}[t]
\includegraphics[angle=0.0,width=1.8in]{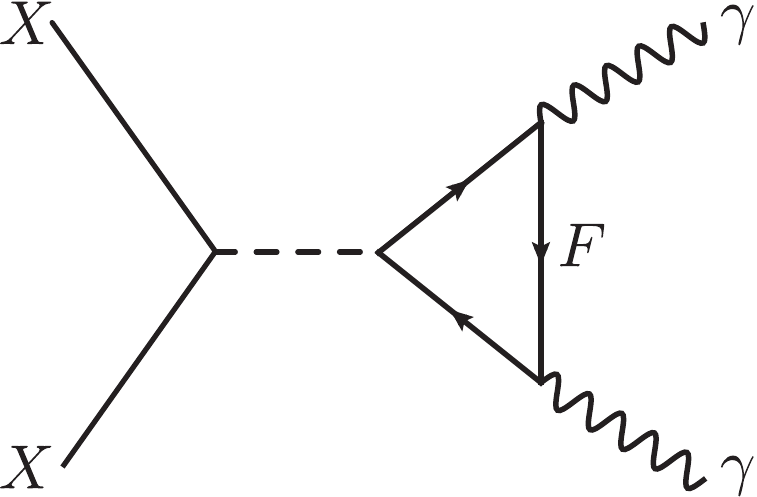}
\caption{An example of a resonance-type diagram for dark matter annihilation to two photons.}
\label{resonance}
\end{figure}

The first case we consider are diagrams in which $s$-channel type annihilation is combined with a charged particle loop (such as that shown in Fig.~\ref{resonance})~\cite{Bergstrom:1997fh}. For this class of diagrams with a scalar mediator, the low-velocity annihilation cross section for two Majorana fermions, $X$, is given by:
\begin{eqnarray}
\sigma v_{\gamma \gamma} &=& \frac{\alpha^2 g^2_{F} g^2_{X}}{256\pi^3}\frac{m^2_F}{[(4m^2_{X}-m^2_M)^2+m^2_M \Gamma^2_M]} \nonumber \\
&\times& \bigg[\arctan[((m^2_F-m^2_X)/m^2_{X})^{-1/2}]  \bigg]^2,
\end{eqnarray}
where $g_{F}$ and $g_{X}$ denote the couplings between the $s$-channel mediator and the charged particle in the loop, and between the mediator and the dark matter, respectively. Here $m_M$ and $m_F$ denote the masses of the $s$-channel mediator and the particle in the loop, respectively. We have assumed that the particle in the loop carries unit charge. Once the condition $m_F \ge m_X$ is met, the cross section to $\gamma \gamma$ is maximized for $m_M=2m_{X}=260$ GeV and $m_F \approx m_X$. In the limit $m_F\rightarrow m_X$, this leads to: 
\begin{eqnarray}
\sigma v_{\gamma \gamma, {\rm max}} &=& \frac{\alpha^2 g^2_{F} g^2_{X}}{4096 \pi\,\Gamma^2_M}   \\
&\approx& 2 \times 10^{-27} {\rm cm}^3/{\rm s} \, \bigg(\frac{g_{F} g_{X}}{1}\bigg)^2 \, \bigg(\frac{5 \, {\rm GeV}}{\Gamma_{M}}\bigg)^2.\nonumber
\end{eqnarray}
Thus such a resonance could potentially generate a cross section large enough to account for the observed 130 GeV feature ($\sigma v_{\gamma \gamma} \sim 2\times 10^{-27}$ cm$^3$/s), although only if each of the following conditions are met: 1) the charged particle is only modestly heavier than the dark matter, 2) the couplings of the mediator to the dark matter and to the charged fermion are large (of order unity), and 3) the resonance lies within $\sim$5 GeV or so of $2 m_{X}$, constituting a $\sim$$2\%$ tuning.

\begin{figure*}[t]
\includegraphics[angle=0.0,width=3.8in]{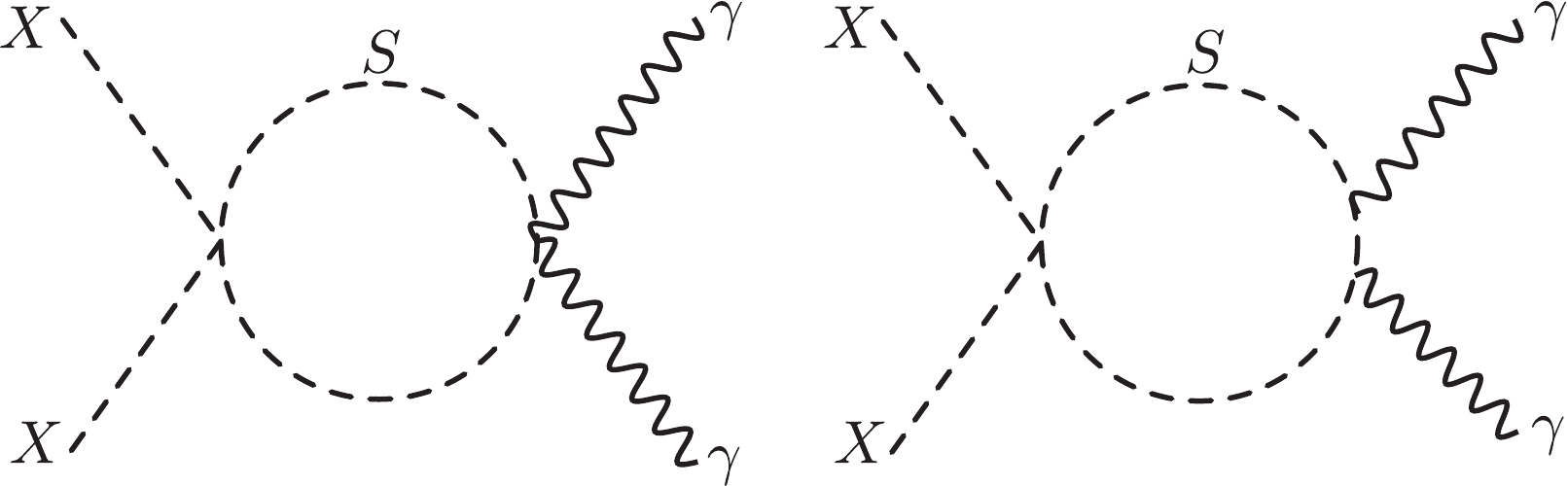}
\caption{A scalar contact interaction for dark matter annihilation to two photons.}
\label{scalar}
\end{figure*}

\subsection{Scalar contact interaction diagrams}

As a second case, we consider scalar dark matter particles which possess an interaction with a charged scalar, $S$, of the form $\mathcal{L}=(\lambda_{X}/2) X^2 |S|^2$~\cite{Cline:2012nw} (see Fig.~\ref{scalar}). Unlike the other cases considered in this section, in order for the $XXSS$ coupling to be renormalizable, neither $X$ nor $S$ can be fermionic. This interaction leads to a low-velocity annihilation cross section to two photons that is given by:
\begin{eqnarray}
\sigma v_{\gamma \gamma} = \frac{\alpha^2 \lambda^2_{X}}{128\pi^3 m^2_X}[1-(m^2_S/m^2_X)\arcsin^2(m_X/m_S)]^2.
\end{eqnarray}
As in the previous subsection, we are forced to consider the case in which $m_S \ge m_X$. For the maximal case (in the limit of $m_S \rightarrow m_X$), this cross section reduces to:
\begin{eqnarray}
\sigma v_{\gamma \gamma,{\rm max}} &=& \frac{\alpha^2 \lambda^2_{X}}{128\pi^3 m^2_X} \, [(\pi^2/4)-1]^2 \\
&\approx& 2.0 \times 10^{-29} \, {\rm cm}^3/{\rm s} \times \bigg(\frac{\lambda_{X}}{1}\bigg)^2 \, \bigg(\frac{130\, {\rm GeV}}{m_X}\bigg)^2. \nonumber
\end{eqnarray}
Thus to accommodate the required cross section with this process, both very large couplings ($\lambda_{X} \sim 10$), and a relatively light charged scalar are necessary.

\subsection{Box-type diagrams}

\begin{figure}[t]
\includegraphics[angle=0.0,width=2.3in]{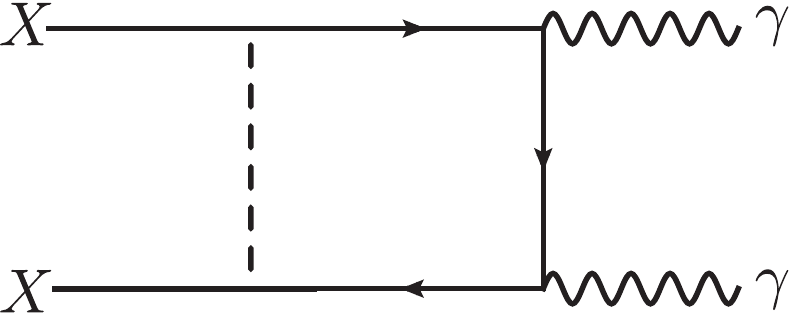}
\caption{An example of a box-type diagram for dark matter annihilation to two photons.}
\label{box}
\end{figure}

The next case we consider are processes in which charged particles are exchanged around a box-like loop (see Fig.~\ref{box})~\cite{Bergstrom:1997fh,Bertone:2009cb}. Within the context of supersymmetry, such diagrams include fermion-sfermion or chargino-charged Higgs loops. While the annihilation cross section for such processes are in general somewhat more difficult to represent analytically, they become much simpler in the maximal allowed limit of $m_F \rightarrow m_X$ (for simplicity, we assume that all of the charged particles in the loop are of the same mass). In this case, such diagrams yield:
\begin{eqnarray}
\sigma v_{\gamma \gamma,{\rm max}} &=& \frac{\alpha^2 g^4_{X}}{4\pi^3 m^2_X} \, [(\pi^2/4)-1]^2 \\
&\approx& 6.4 \times 10^{-28} {\rm cm}^3/{\rm s} \, \bigg(\frac{g_{X}}{1}\bigg)^4 \, \bigg(\frac{130 \, {\rm GeV}}{m_{\chi}}\bigg)^2,\nonumber
\end{eqnarray}
where $g_X$ is the coupling of the dark matter to the particles in the charged loop. Thus once again we find that to accommodate the required large cross section to $\gamma \gamma$, we are forced to consider models with very large couplings ($g_X \gsim 1.4$) and relatively light charged particles.

\subsection{Annihilations to photons through an intermediate state}

The three previous cases have each had a great deal in common. In particular, they each produce their two photon final state through diagrams involving a charged particle loop. As an alternative way of potentially generating the observed gamma-ray feature, we can also consider models in which the dark matter annihilates to a pair of intermediate states (which we will refer to as the dark pion, $\pi_D$), which later decay into pairs of photons. For annihilations to light intermediate states ($\pi_D << m_X$), the gamma-ray spectrum that results from this process is too broad to provide a good fit to the spectral feature reported to be observed in the Fermi data (see the right frame of Fig.~4 in Ref.~\cite{Tempel:2012ey}). If the intermediate state is of comparable mass to the dark matter, however, these particles will be produced approximately at rest, leading to a line-like feature at $E_{\gamma} \sim m_X/2$~\cite{Ibarra:2012dw}.

\begin{figure}[t]
\includegraphics[angle=0.0,width=1.5in]{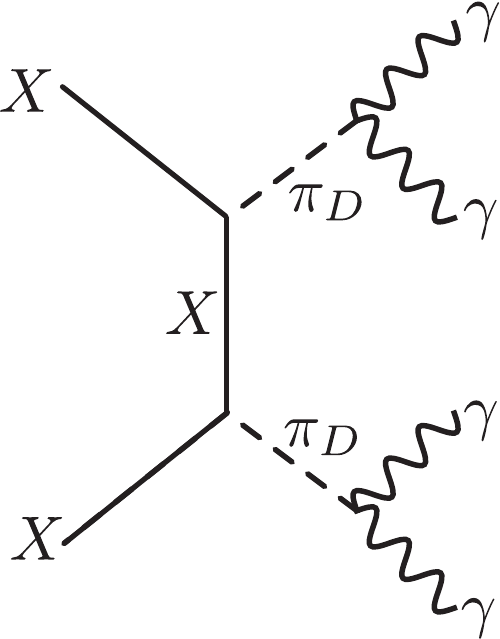}
\caption{Dark matter annihilations to intermediate states which then decay to photons.}
\label{darkpion}
\end{figure}

The cross section for this process (as shown in Fig.~\ref{darkpion}) is given by:
\begin{eqnarray}
\sigma v_{\gamma \gamma} &=& \frac{g^4_{X}}{16 \pi m^2_X} \, \bigg[1-\bigg(\frac{m_{\pi_D}}{m_X}\bigg)^2\bigg]^{1/2} \\
&\approx& 5.5 \times 10^{-27} {\rm cm}^3/{\rm s} \, \bigg(\frac{g_{X}}{0.2}\bigg)^4 \, \bigg(\frac{260 \, {\rm GeV}}{m_{\chi}}\bigg)^2 \nonumber \\
&\times& \bigg[1-\bigg(\frac{m_{\pi_D}}{m_X}\bigg)^2\bigg]^{1/2}. \nonumber
\end{eqnarray}
Here, $g_X$ denotes the coupling of the dark matter to the dark pion, and $m_{\pi_D}$ is the mass of the dark pion. As an example, the parameters $m_{\pi_D} \approx 250$ GeV, $m_X=260$ GeV, and $g_X=0.25$ leads to a cross section of $\sim 4 \times 10^{-27}$ cm$^3$/s, which could account for the observed gamma-ray feature at 130 GeV.

\section{Annihilations to $\gamma Z$ or $\gamma h$}

In addition to processes which generate two photons, dark matter particles in many models can also annihilate to final states consisting of a photon and an additional heavy neutral particle, such as a $Z$~\cite{Ullio:1997ke,bergstromkaplan} or a Higgs boson~\cite{Jackson:2009kg}. Such processes also lead to mono-energetic gamma-ray signatures, although at somewhat lower energies than result in the $\gamma \gamma$ case: $E_{\gamma}=m_X - (m^2_Z/4m_X)$ and $E_{\gamma}=m_X - (m^2_h/4m_X)$, respectively.  In models in which the observed gamma-ray line is produced by dark matter annihilating to $\gamma \gamma$, annihilations to  $\gamma Z$ and/or $\gamma h$ could lead to the appearance of additional gamma-ray lines at energies below 130 GeV. Alternatively, if the dark matter is slightly heavier than 130 GeV and annihilates more often to $\gamma Z$ or $\gamma h$ than to $\gamma \gamma$, such annihilations could potentially account for the line at 130 GeV itself. For an effective field theory approach to dark matter annihilations producing multiple gamma-ray lines, see Ref.~\cite{Rajaraman:2012db}.

For diagrams of the types shown in Figs.~1-3, we can in each case replace one of the final state photons with either a $Z$ or a Higgs boson. The relative brightness of the gamma-ray lines generated from each of the $\gamma \gamma$, $\gamma Z$ and $\gamma h$ final states depend on the spins of the particles involved, and on the couplings between the charged particle in the loop and the photon, $Z$, and Higgs boson. For dark matter particles which annihilate to $\gamma \gamma$, annihilations to $\gamma Z$ are all but inevitable, although the precise ratio of the cross sections for these two line processes depends on the electric charge and weak isospin of the particle in the loop. In many of the most often studied examples, the cross sections to $\gamma \gamma$ and $\gamma Z$ are comparable in magnitude (compare, for example, the results of Refs.~\cite{Bergstrom:1997fh} and~\cite{Ullio:1997ke}). In the specific case of annihilations through a resonant-type diagram (see Fig.~1) with a vector mediator, however, the $\gamma \gamma$ final state is forbidden by the Landau-Yang theorem~\cite{landauyang}. In this case, gamma-ray lines can only appear as a consequence of annihilations to $\gamma Z$ or $\gamma h$.

Annihilations to $\gamma h$ are important in a fairly narrow class of dark matter models. In particular, dark matter in the form of either a scalar or a Majorana fermion will annihilate to $\gamma h$ with a cross section that is highly suppressed in the low-velocity limit relevant for annihilations within the halo of the Milky Way. If the dark matter consists of Dirac fermions or vectors, however, annihilations to $\gamma h$ could be significant, although only if the charged particle in the loop possesses a large coupling to the Higgs boson. Due to its large Yukawa coupling, the top quark makes for an attractive candidate to occupy this loop~\cite{Jackson:2009kg}. The large coupling and multiple colors of the top quark, however, are offset somewhat by its smaller electric charge (2/3), which suppresses the $\gamma-t-t$ vertex.  Also, annihilation through a top quark loop necessarily require a comparable rate of annihilations to $b\bar{b}$ through a top quark-$W^{\pm}$ loop.

If the 130 GeV line under consideration is the result of dark matter annihilating to $\gamma \gamma$, then a second line from annihilations to $\gamma Z$ should be expected to appear at approximately 115 GeV.  The splitting between these two lines is comparable to the energy resolution of Fermi and could likely be distinguished with enough data. If annihilations to $\gamma h$ also occur, then a third line would appear at $\sim$100 GeV (for $m_h\approx$ 125 GeV). The observation of either or both of these additional lines in future data would provide strong supporting evidence in favor of a dark matter interpretation of this feature. If no additional line (at least from $\gamma Z$) is found, it would significantly constrain the range of dark matter models which could account for this signal.

\section{Constraints From The Galactic Center}
\label{constraints}


The most often cited constraints on the dark matter annihilation cross section are those derived from Fermi's observations of dwarf spheroidal galaxies~\cite{dwarf,dwarf2}. The fact that the dark matter distributions in these objects can be directly constrained by their observed stellar dynamics make the resulting constraints quite robust and only modestly dependent on the choice of halo profile model that is adopted.\footnote{While Refs.~\cite{dwarf} and~\cite{dwarf2} each implicitly assume that the dark matter in dwarf spheroidal galaxies follows an NFW profile, these results are based on observations of regions larger than the scale radius of the halo, and thus other choices for the halo profile consistent with the observed stellar kinematics will yield similar constraints.} For the situation at hand, however, the dark matter profile of the Milky Way is constrained by the spatial distribution of the line feature itself, alleviating uncertainties in the constraints derived from annihilations in the the Milky Way's halo. As a result, we can safely apply constraints from gamma-ray observations of the Galactic Center which are significantly more stringent than those derived from dwarf spheroidals. 

The distribution of dark matter in the Milky Way that is required to produce the observed gamma-ray line signal is consistent with that expected from simulations. In particular, Ref.~\cite{Bringmann:2012vr} finds that the distribution of the line-like feature on the sky is best fit by a dark matter distribution which takes the form of a generalized NFW profile with $\gamma=1.1$:
\begin{equation}
\rho(r) = \frac{\rho_0}{(r/R_s)^{\gamma}(1+r/R_s)^{3-\gamma}},
\end{equation}
where $R_s=20$ kpc is the scale radius of the halo and $\rho_0$ is normalized such that $\rho(8.5 \, {\rm kpc})=0.4$ GeV/cm$^3$. Ref.~\cite{Weniger:2012tx} finds that the Fermi data can be approximately equally well fit by a generalized NFW profile with $\gamma=1.15$, or by an Einasto profile:
\begin{equation}
\rho(r)=\rho_0 \exp\bigg(-\frac{2}{\alpha_E}\frac{r^{\alpha_E}}{R_s^{\alpha_E}}\bigg),
\end{equation}
where $\alpha_E=0.17$, $R_s=20$ kpc, and again $\rho_0$ is normalized such that $\rho(8.5 \, {\rm kpc})=0.4$ GeV/cm$^3$.

When we fix the dark matter distribution to that described by these halo models, we can use Fermi's observations of the Galactic Center to place very stringent constraints on the overall dark matter annihilation cross section (to various final states). In particular, following Ref.~\cite{Hooper:2011ti}, we find that for $m_X=130$ GeV, and for the generalized NFW profile described above with $\gamma=1.15$, these observations require that:\footnote{These limits are derived from the Fermi data from the Inner Galaxy, after subtracting known point sources and emission tracing the Galactic Disk. Although one may worry about the possibility of oversubtracting these astrophysical backgrounds, the angular morphologies of these backgrounds are very different from those predicted from dark matter annihilations. Such an oversubtraction would require that the gamma-ray emission from the Galactic Disk drops precipitously and unexpectedly in the inner degree or so around the direction of the Galactic Center, far more than this emission varies elsewhere within $l = \pm 20^{\circ}$. See Ref.~\cite{Hooper:2011ti} for more details.} 
\begin{eqnarray}
\sigma v_{b\bar{b}}&<&8\times10^{-27} {\rm cm}^3/{\rm s}, \\
\sigma v_{W^+ W^-}&<&1.0\times10^{-26} {\rm cm}^3/{\rm s},  \nonumber \\
\sigma v_{ZZ}&<&1.1\times10^{-26} {\rm cm}^3/{\rm s}, \nonumber \\
\sigma v_{c\bar{c}}&<&6\times10^{-27} {\rm cm}^3/{\rm s},\nonumber  \\
\sigma v_{\tau^+ \tau^-}&<&7\times10^{-27} {\rm cm}^3/{\rm s},\nonumber \\
\sigma v_{\mu^+ \mu^-}&<&7\times10^{-26} {\rm cm}^3/{\rm s}. \nonumber
\end{eqnarray}
The constraints derived in the case of an Einasto profile are very similar (about 10\% less stringent) to those found in the $\gamma=1.15$ NFW case.\footnote{The halo profile models considered here are not identical to those considered in Ref.~\cite{Hooper:2011ti}. To derive these limits, we reanalyzed the data used in that study. Also note that the dark matter distributions considered in Ref.~\cite{Hooper:2011ti} are the equivalent of generalized NFW profiles with $R_s\rightarrow \infty$. If that analysis had been performed with $R_s=20$ kpc, annihilation rates approximately three times larger (and constraints approximately three times more strignent) would have been found.} Note that these constraints are appproximately an order of magntiude more stringent than those derived from dwarf spheroidals~\cite{dwarf,dwarf2}. 

These limits on the dark matter's annihilation cross section impose significant constraints on the origin of 130 GeV dark matter particles in the early universe. In particular, if the dark matter is to avoid being overproduced in the early universe, it must annihilate during the freeze-out epoch with a thermally averaged cross section of $\langle \sigma v \rangle_{\rm FO} \ge 3 \times 10^{-26}$ cm$^3$/s. To reconcile this with the constraints listed above, this particle must either annihilate through velocity-suppressed ($p$-wave) processes, or annihilate primarily to final states which do not include quarks, gauge bosons, or taus. If these conditions had not been met, the gamma-ray flux from the Galactic Center would be several times brighter than is observed. Alternatively, one could normalize the dark matter density profiles to a local density of 0.20-0.25 GeV/cm$^3$ (rather than 0.4 GeV/cm$^3$), and thus sufficiently weaken the Galactic Center constraint; although this also requires the annihilation cross section to $\gamma \gamma$ to be larger than the values previously considered by a factor of $\sim$2-4. Lastly, we mention that if the dark matter was never in thermal equilibrium in the early universe, such as in a scenario with a very low ($\sim$10 GeV) reheating temperature, smaller total annihilation cross sections may be possible without resulting in the overproduction of dark matter.

\section{Constraints and Prospects From Colliders}

Due to dark matter's feeble couplings, colliders are in many cases not particularly sensitive to such particles. The most stringent collider constrains on dark matter are generally derived from their ability to produce colored or charged states that appear along with the dark matter candidate within a given theory, rather than by producing dark matter particles themselves. Although searches at the Large Hadron Collider (LHC) for the pair production of dark matter particles in conjunction with a jet or photon also appear promising~\cite{mono}, the constraints from these searches (and from similar searches at the Tevatron) do not yet strongly restrict the viable dark matter parameter space~\cite{Chatrchyan:2012te}.

Although searches for dark matter particles at the LHC can be challenging, many of the scenarios we have discussed in this article require the presence of a relatively light charged particle in order to generate the desired annihilation cross section of dark matter particles to $\gamma \gamma$. In particular, unless the couplings involved are significantly larger than unity or there exists a narrow resonance at a mass within a few percent of twice the dark matter's mass, then the charged particles involved in the relevant diagrams (Figs.~1-3) must fall in the range of $\sim$130-200 GeV. While there exist robust constraints on charged particles lighter than $\sim$100 GeV from LEP II~\cite{pdg}, this mass range required to generate such a gamma-ray line is currently not constrained by colliders.

At the present time, data from the LHC is at best only able to reveal hints of new charged particles in this mass range decaying quickly to neutral massive states~\cite{Lisanti:2011cj} and Standard Model charged particles. However, this window is expected to be quite throughly explored as the LHC collects more data (and eventually achieves higher energy) throughout the remainder of this year and beyond. These charged particles cannot be long-lived (on detector timescales), as such particles are conclusively ruled out by the collider experiments \cite{Aad:2011mb}. However, if the charged and neutral particles were separated by a small mass splitting on the order of 100~MeV-GeV, then resulting lifetime ($c\tau \sim 1$~cm), would make discovery at the LHC extremely challenging, independent of luminosity \cite{FileviezPerez:2008bj}.

Alternatively, several of the scenarios described in this article are expected to lead to the production of events at the LHC which contain multiple photons and missing energy~\cite{Cline:2012nw}. With this signature in mind, analyses such as those described in Ref.~\cite{Aad:2011zj} may be sensitive to many of the dark matter scenarios predicting large annihilation cross sections to gamma-ray lines.

The most difficult model predicting a bright gamma-ray line to test at the LHC is that in which the dark matter annihilates to a pair of heavy neutral particles which subsequently decay to photons (see Sec.~IID and Fig.~4). As in many such hidden sector models, the couplings between the dark pions and their photon decay products could potentially be quite small, making collider searches for new physics of this type very challenging.

\section{Summary and Conclusions}

In this article, we have discussed the implications for particle dark matter of a 130 GeV gamma-ray line, as tentatively identified within the data of the Fermi Gamma-Ray Space Telescope. In particular, we have identified several classes of models which could potentially generate such a signal. Most of these scenarios involve a heavy charged particle loop (see Secs.~IIa, IIb, and IIc). In each of these cases, we find that in order to generate the required cross section to $\gamma \gamma$ (or $\gamma Z$, $\gamma h$), a combination of relatively light ($\sim$130-200 GeV) new charged particles, quite large couplings, and/or very favorable resonances are required. While the couplings needed to accommodate the large cross section to gamma-ray lines are of order $\sim$1-3 and thus are not yet in the non-perturbative regime, such values require that a UV-completion of the theory must appear at a relatively low-scale. One way to relax this requirement would be to consider multiple particles (or particles with multiple ``colors'') which run through these loops, or particles with more than unit electric charge~\cite{Cline:2012nw}. In the later case, one needs to consider how such a particle decays and take care to not violate limits on long-lived charged particles.

Lastly, we note that a scenario in which $m_X \sim260$ GeV dark matter particles annihilate to a pair of heavy neutral states which subsequently decay to photons could also generate a gamma-ray line with the features reported to be present in the Fermi data. Such a model could produce the observed line without the requirement of very large couplings, and could plausibly evade detection at the LHC.

\smallskip

We would like to thank Torsten Bringmann, Paddy Fox, Roni Harnick, Tim Linden and Stefano Profumo for valuable discussions. The authors are supported by the US Department of Energy.

\end{document}